\documentclass[pra, aps, showpacs, reprint]{revtex4-1}

\usepackage[utf8]{inputenc}
\usepackage{graphicx}
\usepackage{amsmath}
\usepackage{amssymb}
\usepackage{braket}
\usepackage{color}
\usepackage{tikz}
\usetikzlibrary{positioning}
\usepackage{pdfpages}

\makeatletter
\AtBeginDocument{\let\LS@rot\@undefined}
\makeatother

\newcommand{\diff}[1]{\mathrm{d} #1}

\newcommand{\ve}[1]{\boldsymbol{\mathbf{#1}}}

\providecommand{\abs}[1]{\lvert#1\rvert} 

\definecolor{darkgreen}{RGB}{0, 150, 0}
\definecolor{cyan2}{RGB}{0, 255, 255}

\begin{document}

\author{{\O}yvind Johansen}
\email{Corresponding author: oyvinjoh@ntnu.no}
\author{Arne Brataas}
\affiliation{Center for Quantum Spintronics, Department of Physics, Norwegian University of Science and Technology, NO-7491 Trondheim, Norway}

\date{\today}
\title
{Non-local coupling between antiferromagnets and ferromagnets in cavities}

\begin{abstract}
Microwaves couple to magnetic moments in both ferromagnets and antiferromagnets. Although the magnons in ferromagnets and antiferromagnets radically differ, they can become entangled via strong coupling to the same microwave mode in a cavity. The equilibrium configuration of the magnetic moments crucially governs the coupling between the different magnons because the antiferromagnetic and ferromagnetic magnons have the opposite spins when their dispersion relations cross. We derive analytical expressions for the coupling strengths and find that the coupling between antiferromagnets and ferromagnets is comparable to the coupling between two ferromagnets. Our findings reveal a robust link between cavity spintronics with ferromagnets and antiferromagnets.
\end{abstract}

\maketitle

Magnets and photons can couple strongly and coherently on the quantum level \cite{Soykal:prl:2010}.
This coupling has been observed as hybridizations between ferromagnetic magnons and either microwave resonators \cite{Huebl:prl:2013}, microwave photons \cite{Tabuchi:prl:2014}, or optical photons \cite{Osada:prl:2016,Zhang:prl:2016,Haigh:prl:2016,Braggio:prl:2017}.
Together, these findings constitute the birth of \textit{cavity spintronics}, a new interdisciplinary field with roots in spintronics, cavity quantum electrodynamics and quantum optics. Shortly after the initial observations of magnon-photon hybridization, experiments also achieved ultrastrong coupling between magnons and microwaves \cite{Zhang:prl:2014,Goryachev:pra:2014,Bourhill:prb:2016,Kostylev:apl:2016}. One can tune the coupling strength by changing the temperature of the system \cite{Maier-Flaig:apl:2017,Boventer2018}. The cooperativity of the hybridization, namely, the ratio of the coupling strength to the loss rates, is a measure of the coherence in the system and can be as large as $10^7$ \cite{Bourhill:prb:2016}. The transmission and reflection coefficients of a cavity are measures of the magnon-photon coupling.  Brillouin light scattering \cite{Klingler:apl:2016} or spin pumping  \cite{Cao:prb:2015,Bai:prl:2015,Bai:prl:2017} can provide additional information. Although the focus thus far has been placed on the hybridization between ferromagnetic magnons and photons, theory predicts that there should also be a significant coupling between antiferromagnetic magnons and photons \cite{Yuan:apl:2017}. Recently, a robust coupling between microwave photons and antiferromagnetic fluctuations in an organic magnet has been observed \cite{Mergenthaler:prl:2017}.

When two ferromagnets or ferrimagnets couple to the same cavity mode or the same photons, a non-local interaction between the magnons emerges \cite{Lambert:pra:2016,Rameshti:prb:2018}. The quantized magnetic field of the cavity mode or photons mediates this indirect coupling. In turn, this coupling facilitates the coherent transport of magnons over macroscopic distances. 

The magnons in ferromagnets and antiferromagnets strongly differ. Nevertheless, we will demonstrate that a clear non-local coupling can also arise between the magnons in an antiferromagnet and those in a ferromagnet.
This link opens the door toward long-range spin communication between antiferromagnetic and ferromagnetic spintronic devices. Such a connection has so far been elusive because the vanishing magnetization of the antiferromagnet renders it invisible to dipolar interactions with the ferromagnet. The need for and interest in such a long-range coupling between dissimilar magnetic materials have increased with the recent emergence of antiferromagnetic spintronics \cite{Cheng:prl2014,Marti2014,Cheng:prl2016,Jungwirth2016,Baltz2018}. Our findings open a path toward integrating these new components with existing ferromagnetic ones.

\begin{figure}[h!]
\centering
{\includegraphics[width=\linewidth]{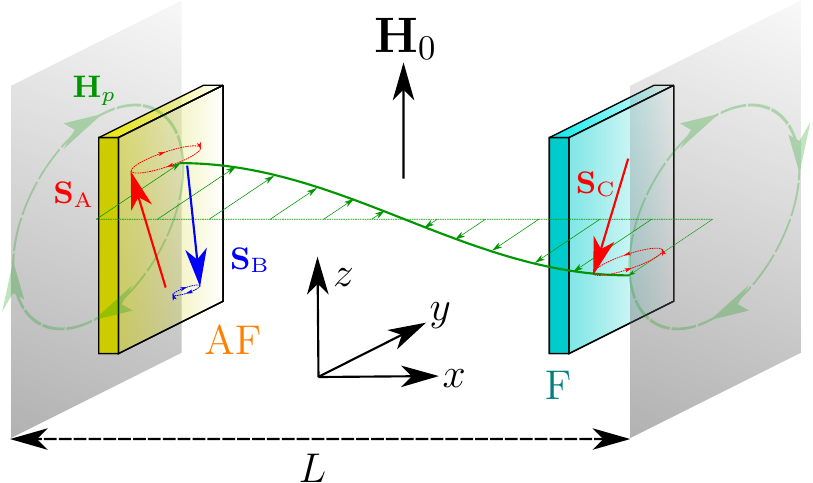}}
\caption{A microwave cavity consisting of two perfect conductors separated by a distance $L$. The magnons in an antiferromagnet (with sublattice spins $\ve{S}_\text{A}$ and $\ve{S}_\text{B}$) and those in a ferromagnet (with spins $\ve{S}_\text{C}$) couple via the quantized magnetic field $\ve{H}_p$. At equilibrium, the magnetic moments are collinear along the $z$ axis. Detection of the coupling is possible when the frequencies of the magnons are similar. This can be achieved by applying an external magnetic field $\ve{H}_0$ along the equilibrium axis of the magnetic moments. }
\label{fig:Cavity}
\end{figure}

To study the cavity-mediated coupling between an antiferromagnet and a ferromagnet, we consider the setup depicted in Fig. \ref{fig:Cavity}. The Hamiltonian consists of four components:
\begin{equation}
H=H_\text{AF}+H_\text{F}+H_\text{ph}+H_\text{m-ph} \, ,
\label{eq:FullH}
\end{equation}
where $H_\text{AF}$ describes the antiferromagnet; $H_\text{F}$, the ferromagnet; $H_\text{ph}$, the cavity modes; and $H_\text{m-ph}$, the coupling of the magnons in both the antiferromagnet and the ferromagnet to the cavity modes.
We consider an antiferromagnet described by the Hamiltonian
\begin{align}
\nonumber H_\text{AF} &= J \sum_{\langle i,j\rangle \in \text{AF}}\ve{S}_i\cdot\ve{S}_j + \abs{\gamma}\sum_{i\in \text{AF}}\ve{H}_0\cdot\ve{S}_i \\
&-\frac{K_\parallel}{2}\sum_{i\in \text{AF}}\left(S_i^z\right)^2+\frac{K_\perp}{2}\sum_{i\in \text{AF}}\left(S_i^x\right)^2 \, ,
\label{eq:HAF}
\end{align}
where $J>0$ is the exchange coupling between the spins $\ve{S}$, $\gamma$ is the gyromagnetic ratio, $\ve{H}_0=H_0\ve{\hat{z}}$ is an external magnetic field (in units of Tesla) along the easy axis, and $K_\parallel>0$ and $K_\perp \geq 0$ are the easy- and hard-axis anisotropy constants, respectively.
For the ferromagnet, we consider energy contributions from the Zeeman and dipolar interactions.
It will become clear that the photons couple only to homogeneous magnetic excitations. 
We can therefore use a simplified expression for the dipolar interaction, resulting in a ferromagnetic Hamiltonian of
\begin{align}
H_\text{F} = \abs{\gamma}\sum_{i\in \text{F}}\ve{H}_0\cdot\ve{S}_i + \frac{\mu_0}{2}\int_{V_\text{F}}\diff V \sum_{i=x,y,z} N_i M_i^2 \, .
\end{align}
Here, $N_{x,y,z}$ are the elements of the diagonal demagnetization tensor, $V_\text{F}$ is the volume of the ferromagnet, $\mu_0$ is the vacuum permeability, and $M_i$ is the $i$-th component of the magnetization. 
The magnetization is related to the spins by $\ve{M}=-\abs{\gamma}\sum_{i\in \text{F}}\ve{S}_i/V_\text{F}$. 

Now, let us rewrite the two magnetic Hamiltonians in terms of magnon operators. We perform a Holstein-Primakoff transformation \cite{HolsteinPrimakoff} of the spin operators $\ve{S}_i$, retaining terms of up to second order in the magnon operators in the resulting Hamiltonian. Following this procedure, we perform a Fourier transformation from spatial coordinates into a momentum representation. To diagonalize the antiferromagnetic and ferromagnetic Hamiltonians, we also perform Bogoliubov transformations. 

For the antiferromagnet, the Bogoliubov transformation is four-dimensional in the general case, when ${K_\perp\neq 0}$. 
In terms of the initial Holstein-Primakoff magnon annihilation (creation) operators $a_{\ve{q}}^{(\dagger)}$ and $b_{\ve{q}}^{(\dagger)}$ within sublattices A and B, respectively (see Fig. \ref{fig:Cavity}), the Bogoliubov transformation into the diagonal eigenmode magnons $\alpha_{\ve{q}}^{(\dagger)}$ and $\beta_{\ve{q}}^{(\dagger)}$ is \cite{Kamra:prb:2017}
\begin{align}
\begin{pmatrix}
\alpha_{\ve{q}} \\ \beta_{-\ve{q}}^\dagger \\ \alpha_{-\ve{q}}^\dagger \\ \beta_{\ve{q}}
\end{pmatrix}
=
\begin{pmatrix}
u_{\alpha,a} & v_{\alpha,b} & v_{\alpha,a} & u_{\alpha,b} \\
v_{\beta,a}^* & u_{\beta,b}^* & u_{\beta,a}^* & v_{\beta,b}^* \\
v_{\alpha,a}^* & u_{\alpha,b}^* & u_{\alpha,a}^* & v_{\alpha,b}^* \\
u_{\beta,a} & v_{\beta,b} & v_{\beta,a} & u_{\beta,b}
\end{pmatrix}
\begin{pmatrix}
a_{\ve{q}} \\ b_{-\ve{q}}^\dagger \\ a_{-\ve{q}}^\dagger \\ b_{\ve{q}}
\end{pmatrix} \, .
\end{align}
Here, $\ve{q}$ is the wave vector of the magnons.
To determine the elements of the Bogoliubov transformation, we impose the requirement that the operators $\alpha_{\ve{q}}^{(\dagger)}$ and $\beta_{\ve{q}}^{(\dagger)}$ must satisfy bosonic commutation relations as well as the relations $\left[\alpha_{\ve{q}},H_\text{AF}\right]=\hbar\omega_\alpha\alpha_{\ve{q}}$ and $\left[\beta_{\ve{q}},H_\text{AF}\right]=\hbar\omega_\beta\beta_{\ve{q}}$. 
Here, $\omega_{\alpha,\beta}$ are the eigenfrequencies of the antiferromagnet. The excitations in the antiferromagnet are then
\begin{equation}
H_\text{AF}=\sum_{\ve{q}} \left(\hbar\omega_\alpha \alpha_{\ve{q}}^\dagger \alpha_{\ve{q}}+\hbar\omega_\beta \beta_{\ve{q}}^\dagger \beta_{\ve{q}}\right) \, .
\end{equation}
Let us define the frequencies $\omega_E=\hbar J s Z$, $\omega_H=\abs{\gamma} H_0$, $\omega_\parallel=\hbar s K_\parallel$, and $\omega_\perp=\hbar s K_\perp$, where $s$ is the spin number and $Z$ is the number of nearest neighbors in the antiferromagnet; then, we can express the eigenfrequencies as 
\begin{align}
\nonumber \omega_{\pm}^2 &\approx \omega_E\left(2\omega_\parallel+\omega_\perp\right)+\omega_H^2 \\
&\pm \sqrt{\omega_E^2\omega_\perp^2+4\omega_H^2\omega_E\left(2\omega_\parallel+\omega_\perp\right)} \, ,
\end{align}
where $\omega_\alpha=\omega_+$ and $\omega_\beta=\omega_-$. 
For simplicity, we have here included only terms to the lowest order in the anisotropy frequencies  $\omega_{\parallel / \perp}$, as these are typically much smaller than the exchange frequency $\omega_E$. We also assume that the wavelength of the magnons is much larger than the atomic length scale.

In the long-wavelength limit ($\ve{q}=\ve{0}$), the quantized ferromagnetic Hamiltonian can be expressed as \cite{Kamra:prl:2016,Kamra:prb:2017}
\begin{align}
H_\text{F}&=A_\text{F} c^\dagger c+B_\text{F}\left(c c+c^\dagger c^\dagger\right) \, ,
\end{align}
where ${A_\text{F}=\hbar\left[\omega_H+\omega_M\left(N_{xz}+N_{yz}\right)/2\right]}$ and ${B_\text{F}=\hbar\omega_M N_{xy}/4}$.
Here, we have defined ${\omega_M=\abs{\gamma} \mu_0 M_s}$, where $M_s$ is the saturation magnetization of the ferromagnet, and ${N_{ij}=N_i-N_j}$.
We have also introduced the ferromagnetic magnon annihilation (creation) operator $c^{(\dagger)}$ from the initial Holstein-Primakoff transformation for the ferromagnet.
This Hamiltonian can be diagonalized through a Bogoliubov transformation defined by $\eta=u_\text{F} c + v_\text{F} c^\dagger$. 
Imposing bosonic commutation relations on $\eta^{(\dagger)}$ yields the constraint $u_\text{F}^2-v_\text{F}^2=1$. 
The elements of the Bogoliubov transformation are \cite{Kamra:prl:2016} 
\begin{subequations}
\begin{align}
v_\text{F} &=2B_\text{F}/\sqrt{\left(A_{\text{F}}+\hbar\omega_{\eta}\right)^2-4B_\text{F}^2} \, , \\ 
u_\text{F} &=v_\text{F}\left(A_{\text{F}}+\hbar\omega_{\eta}\right)/\left(2B_{\text{F}}\right) \, .
\end{align}
\label{eq:Bogoliubov_F}
\end{subequations}
The energy of the $\eta$ magnons is $\hbar \omega_{\eta} = \sqrt{A_\text{F}^2-4B_\text{F}^2}$.
We assume a thin-film geometry with $N_x=1$ and $N_y=N_z=0$. The resonance frequency is then ${\omega_{\eta}=\sqrt{\omega_H\left(\omega_H+\omega_M\right)}}$.

Now, let us consider the cavity. We first introduce a simple geometry. We will later argue that the simple geometry also captures the essential physics in more  complex cavities. We describe the microwave cavity as two perfect conductor plates separated by a distance $L$ in the $x$ direction. The plates are located at $x=0$ and $x=L$. The boundary conditions near the conducting plates cause a quantization of the electromagnetic modes in the $x$ direction. The allowed wave numbers in this direction are $n\pi/L$ ($n=1,2,\ldots$). We assume that the microwaves propagate only in the direction perpendicular to the conducting plates, i.e., along the $x$ axis. The resulting quantized magnetic field is
\begin{align}
\nonumber \ve{H}_p (\ve{r}) &= \sum_{n,\lambda} i\cos\left(\frac{n\pi x}{L}\right)\left(\frac{\hbar\omega_n\mu_0}{V}\right)^{1/2} \\
&\times\left(p_{n,\lambda}\ve{\hat{x}}\times\ve{\hat{e}}_\lambda - p_{n,\lambda}^\dagger\ve{\hat{x}}\times\ve{\hat{e}}_\lambda^* \right) \, .
\label{eq:Hp}
\end{align}
Here, $p_{n,\lambda}^{(\dagger)}$ is the boson annihilation (creation) operator associated with a cavity mode with quantum numbers $n$ and $\lambda$. $\lambda$ denotes the polarization, and $\ve{\hat{e}}_\lambda$ is the polarization unit vector.
$V$ is the volume of the cavity, $\omega_n=n\pi c/L$ is the cavity mode frequency, and $c$ is the speed of light. The cavity Hamiltonian is 
\begin{align}
H_\text{ph}
&=\sum_{n,\lambda} \hbar \omega_n \left(p_{n,\lambda}^\dagger p_{n,\lambda}+\frac{1}{2}\right) \, .
\end{align}
The cavity modes can couple to the magnons in both antiferromagnets \cite{Yuan:apl:2017} and ferromagnets \cite{Huebl:prl:2013,Tabuchi:prl:2014}. The interaction is represented by a Zeeman term between the quantized magnetic field $\ve{H}_p$ in Eq. \eqref{eq:Hp} and the spins in the magnetic materials. 
This interaction is described by the magnon-photon Hamiltonian $H_\text{m-ph}$.

We use a circular polarization basis, $\lambda=\pm$ and ${\ve{\hat{e}}_\lambda=\ve{\hat{e}}_{\pm}=\left(\ve{\hat{z}}\pm i \ve{\hat{y}}\right)/\sqrt{2}}$, for the cavity modes. Since the speed of light is much greater than the group velocity of the magnons, the cavity modes only couple to magnons of approximately $\abs{\ve{q}}=0$.  For simplicity, we assume that the cavity only couples to the $\ve{q}=\ve{0}$ modes. We also use the rotating wave approximation (RWA) to simplify the magnon-photon interaction. In the RWA, rapidly oscillating terms, e.g., pairs of creation operators or pairs of annihilation operators, are disregarded because their average contributions quickly become small.

We tune the system size such that the $n=1$ cavity mode has a frequency near the crossing of the ferromagnetic and antiferromagnetic dispersion relations.
The coupling between the ferromagnet and the antiferromagnet is significant only near this crossing region. Therefore, we disregard any contributions from the high-energy $n>1$ cavity modes and $\alpha$ magnons of the antiferromagnet. There are also three-particle (and higher-order) interactions in which pairs of magnon creation/annihilation operators couple to the cavity mode, i.e., terms such as $\beta^\dagger\beta^\dagger p_{n,\lambda}$ and $\beta\beta p_{n,\lambda}^\dagger$. Such couplings to the $n=2$ cavity mode, for example, oscillate sufficiently slowly to be significant in the RWA.  However, these higher-order contributions scale more weakly with the number of spins in the ferromagnet/antiferromagnet ($N_\text{F/AF}$) than the two-particle interactions do. Such terms can therefore be safely disregarded.
 
With these approximations and simplifications, the magnon-photon interaction Hamiltonian reduces to
\begin{align}
\label{eq:Hmph}
H_\text{m-ph} &= \hbar \left(\tilde{g}_\text{AF}\beta^\dagger-\tilde{g}_\text{F}\eta^\dagger\right)\left(p_++p_-\right) + \text{H.c.} \, .
\end{align}
Here, we have introduced the ferromagnetic coupling strength $\tilde{g}_\text{F}=g_\text{F}\left(u_\text{F}+v_\text{F}\right)$ and the antiferromagnetic coupling strength $\tilde{g}_\text{AF}=g_\text{AF}\left(u_{\beta,a}-v_{\beta,b}+v_{\beta,a}-u_{\beta,b}\right)/\sqrt{2}$, where $g_{\text{F/AF}}=\abs{\gamma}\sqrt{\hbar\omega_{p}\mu_0 s N_{\text{F/AF}}/(4V)}$, with $\omega_p=\omega_{n=1}$ being the microwave frequency. In addition, the coupling strength in a cavity depends on the overlap of the cavity mode with the magnetic materials \cite{Cao:prb:2015}. More general geometries are described  by substituting $g_\text{F/AF}\rightarrow\xi_\text{F/AF}g_\text{F/AF}$, where $\xi_\text{F/AF}$ is a geometrical overlap factor of order one for the ferromagnet/antiferromagnet \cite{Boventer2018}. In the following illustrations, we consider a perfect overlap between the cavity mode and each magnetic material; i.e., we set $\xi_\text{F,AF}=1$.
In general, the elements of the Bogoliubov transformation for the antiferromagnetic magnons ($u_{\beta,a}$, $v_{\beta,b}$, $v_{\beta,a}$, and $u_{\beta,b}$) are complicated functions of the parameters of the antiferromagnetic Hamiltonian expressed in Eq. \eqref{eq:HAF}. We therefore present the coupling strengths between the antiferromagnetic magnons and the cavity for only two particular scenarios. In the first scenario, the antiferromagnet is uniaxial ($\omega_\perp=0$), and we find that the coupling strength is 
\begin{equation}
\frac{\tilde{g}_\text{AF}}{g_\text{AF}}\approx \left(\frac{\omega_\parallel}{8\omega_E}\right)^{1/4} \, .
\label{eq:gAFuni}
\end{equation}
In the second scenario, the antiferromagnet has an easy-plane anisotropy such as that in NiO, where $\mathcal{O}(\omega_\perp)\sim\mathcal{O}(\sqrt{\omega_\parallel \omega_E})$ \cite{Satoh:prl:2010,Hutchings:prb:1972}. The coupling strength is then
\begin{align}
\frac{\tilde{g}_\text{AF}}{g_\text{AF}}\approx \frac{\omega_H}{\sqrt{2\omega_E}\left(2\omega_E\omega_\parallel-\omega_H^2\right)^{1/4}} \, .
\label{eq:gAFep}
\end{align}
The expressions given in (\ref{eq:gAFuni}) and (\ref{eq:gAFep}) are valid to the lowest order in $\omega_\parallel/\omega_E$.

We consider two geometries. In the first case, the magnetic moments of the ferromagnet and antiferromagnet are perpendicular to the propagation direction of the cavity modes, as in Fig. \ref{fig:Cavity}. In this geometry, which is the one we have considered so far, the magnons couple to both circular polarizations with the same strength, as seen in Eq. \eqref{eq:Hmph}. This is because, in the RWA and to leading order in the spin fluctuations, the component of the quantized magnetic field along the magnetic moments does not influence the dynamics of the magnetic moments. Since the magnets experience an oscillating magnetic field along only one axis, they effectively couple to a cavity mode that is linearly polarized.  The coupling of antiferromagnetic and ferromagnetic magnons requires that they couple to the same cavity mode with quantum numbers $n$ and $\lambda$. Linearly polarized microwaves couple to magnons with different spins and can therefore couple equally strongly to both antiferromagnetic and ferromagnetic magnons. Consequently, this geometry enables a magnon-magnon interaction between an antiferromagnet and a ferromagnet.

In the second geometry, the magnetic moments are collinear with the propagation direction of the cavity modes. In such a geometry,  the coupling strengths strongly depend on the cavity mode polarization. Each magnon couples to the microwave mode with the same spin. 
At the crossing, the antiferromagnetic and ferromagnetic magnons therefore couple to different polarization states $\lambda$ of the cavity modes. Cavity modes with different polarizations $\lambda=\pm$ are independent. This geometry therefore does not lead to a magnon-magnon coupling.
This lack of indirect coupling can be circumvented when at least one of the magnon modes is squeezed or hybridized (i.e., it has a non-integer spin \cite{Kamra:prb:2017}). In this case, the magnons have finite, but dissimilar, couplings to both polarizations of the cavity mode.
Although this generates a finite magnon-magnon coupling, this coupling will typically be rather weak because the magnons still primarily couple to different cavity mode polarizations. We therefore focus on the former geometry, illustrated in Fig. \ref{fig:Cavity}. In this geometry, there are equally strong couplings to both polarizations and, consequently, a stronger coupling between the magnons.

More complex cavity structures can be effectively reduced to our planar cavity geometry when there is only  one mode with a frequency near the crossing point of the magnon dispersion relations. 
The magnon-magnon coupling is then dominated by this mode. The overlap factors $\xi_\text{F,AF}$ depends on the details of the cavity structure.

We now wish to determine the coupling strength between the ferromagnetic and antiferromagnetic magnons. 
We consider a dispersive regime, in which the frequency of the cavity mode is slightly detuned from the crossing point between the ferromagnetic and antiferromagnetic magnon dispersion relations. 
We assume that this detuning $\Delta=\omega_p-\omega_\eta=\omega_p-\omega_\beta$ is much larger than the coupling strengths $\tilde{g}_\text{F,AF}$ and that the frequency of the cavity mode is sufficiently close to the crossing point where $\omega_\eta=\omega_\beta$.
We can then determine the magnon-magnon coupling by performing a unitary transformation of the Hamiltonian given in Eq. \eqref{eq:FullH} and then applying second-order degenerate perturbation theory, with $\tilde{g}_\text{F,AF}$ as the perturbation parameters.
We thus compute that at the crossing point between the antiferromagnetic and ferromagnetic magnon dispersion relations, the transformed Hamiltonian describing the excitations is
\begin{align}
\nonumber H' = &\hbar\left[\omega_\beta-\frac{2\tilde{g}_\text{AF}^2}{\Delta}\right]\beta^\dagger\beta + \hbar\left[\omega_\eta-\frac{2\tilde{g}_\text{F}^2}{\Delta}\right]\eta^\dagger\eta  \\
\nonumber &+\hbar\left[\omega_p+\frac{\tilde{g}_\text{F}^2+\tilde{g}_\text{AF}^2}{\Delta}\right]\left(p_-^\dagger p_- + p_+^\dagger p_+ \right) \\
&+2\hbar\frac{\tilde{g}_\text{F}\tilde{g}_\text{AF}}{\Delta}\left(\beta^\dagger\eta + \eta^\dagger\beta\right)
\label{eq:H_magnon-magnon}
\end{align}
to second order in $\tilde{g}_\text{F,AF}$. The first three terms in Eq. \eqref{eq:H_magnon-magnon} describe shifts in the frequencies of both the magnon and cavity modes, whereas the last term describes a dispersive coupling between the ferromagnetic and antiferromagnetic magnons.

It is instructive to find numerical estimates of the effective coupling strengths for different combinations of materials. 
We consider a scenario in which the volume of each of the magnetic materials constitutes 1\% of the volume of the cavity; this magnitude is similar to that in previous experiments involving ferromagnets in spherical cavities \cite{Lambert:pra:2016,Rameshti:prb:2018}.
As ferromagnets, we consider yttrium-iron-garnet (YIG) and Co-Fe alloys. 
YIG is commonly used in cavity spintronics due to its high spin density and incredibly low damping.
We also consider a ferromagnetic Co-Fe alloy because it can have a damping as low as ${\sim 10^{-4}}$ \cite{Schoen2016} and has an even higher spin density than YIG, enabling stronger coupling to the cavity mode. 
The spin density $s N_\text{F}/V_\text{F}$ in YIG is ${\sim 2\cdot 10^{22} \text{ cm}^{-3}}$ \cite{YIGspindensity}. 
In a Co-Fe alloy, it is possible to achieve a spin density as high as ${\sim 2\cdot 10^{23} \text{ cm}^{-3}}$, where we have estimated the magnitude of the spin density as $M_s/\mu_\text{B}$, with $\mu_\text{B}$ being the Bohr magneton. 
The saturation magnetizations of these ferromagnetic materials are $\mu_0 M_s=0.247$ T for YIG \cite{YIGMs} and $\mu_0 M_s\approx 2.4$ T for the Co-Fe alloy \cite{Schoen2016}. 
As antiferromagnets, we consider the uniaxial antiferromagnetic materials MnF$_2$ and NaNiO$_2$ as well as the easy-plane antiferromagnetic material NiO. 
Estimates of the spin densities in MnF$_2$, NaNiO$_2$ and NiO yield ${\sim 10^{22} \text{ cm}^{-3}}$ \cite{Kotthaus:prl:1972}, ${\sim5\cdot 10^{22} \text{ cm}^{-3}}$ \cite{NaNiO2} and ${\sim  10^{23} \text{ cm}^{-3}}$ \cite{Schron:prb:2015}, respectively. 
The exchange and anisotropy frequencies of the antiferromagnetic materials are listed in Table \ref{tab:AFfrequencies}.
\begin{table}[tpb]
	\centering
		\caption{Exchange and anisotropy frequencies of the antiferromagnets.} 
	\begin{tabular}{l l l l} \hline \hline
	Material & $\omega_E$ ($10^{12}$ s$^{-1}$)& $\omega_{\parallel}$ ($ 10^{12}$ s$^{-1}$) & $\omega_{\perp}$ ($10^{12}$ s$^{-1}$)\\ \hline
	MnF$_2$ \cite{Ross:TUMunchen:2013} & 9.3 & 1.5 $\cdot 10^{-1}$ & - \\
	NaNiO$_2$ \cite{NaNiO2} & 8.4$\cdot 10^{-1}$ & 6.2 $\cdot 10^{-2}$ & - \\
 	NiO \cite{Satoh:prl:2010,Hutchings:prb:1972} & 1.7$\cdot 10^{2}$& 2.3 $\cdot 10^{-3}$ & 1.3 $\cdot 10^{-1}$ \\ \hline \hline
	\end{tabular}
	\label{tab:AFfrequencies}
\end{table}
Using these parameters, and choosing the $n=1$ frequency mode of the cavity to lie at the crossing point between the antiferromagnetic and ferromagnetic dispersion relations, we obtain the coupling strengths given in Table \ref{tab:couplingstrengths}.
\begin{table}[tpb]
	\centering
		\caption{Coupling strengths for different combinations of ferromagnets and antiferromagnets.} 
	\begin{tabular}{l c c} \hline \hline
	Materials (AF / F) & $\tilde{g}_\text{AF}/(2\pi)$ (GHz) & $\tilde{g}_\text{F}/(2\pi)$ (GHz) \\ \hline
	MnF$_2$ / YIG & 0.31 & 2.30 \\
	MnF$_2$ / Co-Fe & 0.33 & 7.67 \\
	NaNiO$_2$ / YIG & 0.44 & 1.28 \\
	NaNiO$_2$ / Co-Fe & 0.50 & 5.03 \\
	NiO / YIG & 0.17 & 2.03 \\
	NiO / Co-Fe & 0.15 & 6.88 \\ \hline \hline
	\end{tabular}
	\label{tab:couplingstrengths}
\end{table}

Our results reveal that the coupling strength of the cavity with the antiferromagnet is typically an order of magnitude lower than that with the ferromagnet. This is due to the weak interaction between antiferromagnets and magnetic fields, which gives rise to a suppression factor from the high exchange energy, as seen in Eqs. \eqref{eq:gAFuni} and \eqref{eq:gAFep}.
Importantly, it is still possible to achieve a sizeable magnon-magnon coupling between the antiferromagnet and the ferromagnet. If we detune the cavity from the crossing point of the magnon dispersion relations by $\Delta=5\tilde{g}_\text{F}$, then the magnon-magnon coupling becomes $\approx 2\tilde{g}_\text{AF}/5$.  For MnF$_2$/Co-Fe, this would correspond to a coupling strength of 132 MHz; for NaNiO$_2$/Co-Fe, the corresponding value is 200 MHz, and for NiO/YIG, it is 68 MHz.
These magnitudes are similar to the coupling strength observed between two YIG spheres \cite{Lambert:pra:2016,Rameshti:prb:2018}. 
The inefficiency of the coupling between the antiferromagnet and the cavity is compensated for by the high frequency at the crossing point between the magnon dispersion relations, with a proportionality of $\tilde{g}_{\text{F/AF}}\propto \sqrt{\omega}$. The magnon dispersion relations cross at {148 GHz} {($H_0$=4.2 T)} for MnF$_2$/Co-Fe, at {36 GHz} {($H_0$=0.6 T)} for NaNiO$_2$/Co-Fe, and at {100 GHz} {($H_0=3.5 \text{ T}$)} for NiO/YIG.

Although the linewidths of many ferromagnets have been extensively studied, there are very few similar measurements available for antiferromagnets. 
It is possible that the linewidth may become larger in antiferromagnets since the exchange energy often dominates.
If this is the case, the cooperativity (which is a measure of the coherence within the system) will be lower in systems involving antiferromagnets than in systems consisting solely of ferromagnets. 

In summary, we have shown that despite having opposite spins, antiferromagnetic magnons and ferromagnetic magnons can be coupled inside a microwave cavity. This coupling is strongest when the magnetic moments in the antiferromagnet and ferromagnet are perpendicular to the propagation direction of the cavity modes at equilibrium. In this geometry, the cavity mode couples to both the antiferromagnetic and ferromagnetic magnons, resulting in a robust non-local magnon-magnon coupling. The magnitude of this coupling is similar to that of the non-local interaction between two ferromagnets. 

\textit{Acknowledgments.} The authors would like to thank Mathias Kl\"{a}ui for valuable input. This work was supported by the Research Council of Norway through its Centres of Excellence funding scheme, Project No. 262633 “QuSpin" and Grant No. 239926 “Super Insulator Spintronics,” as well as by the European Research Council via Advanced Grant No. 669442 “Insulatronics.”
\appendix


\begin{thebibliography}{38}%
\makeatletter
\providecommand \@ifxundefined [1]{%
 \@ifx{#1\undefined}
}%
\providecommand \@ifnum [1]{%
 \ifnum #1\expandafter \@firstoftwo
 \else \expandafter \@secondoftwo
 \fi
}%
\providecommand \@ifx [1]{%
 \ifx #1\expandafter \@firstoftwo
 \else \expandafter \@secondoftwo
 \fi
}%
\providecommand \natexlab [1]{#1}%
\providecommand \enquote  [1]{``#1''}%
\providecommand \bibnamefont  [1]{#1}%
\providecommand \bibfnamefont [1]{#1}%
\providecommand \citenamefont [1]{#1}%
\providecommand \href@noop [0]{\@secondoftwo}%
\providecommand \href [0]{\begingroup \@sanitize@url \@href}%
\providecommand \@href[1]{\@@startlink{#1}\@@href}%
\providecommand \@@href[1]{\endgroup#1\@@endlink}%
\providecommand \@sanitize@url [0]{\catcode `\\12\catcode `\$12\catcode
  `\&12\catcode `\#12\catcode `\^12\catcode `\_12\catcode `\%12\relax}%
\providecommand \@@startlink[1]{}%
\providecommand \@@endlink[0]{}%
\providecommand \url  [0]{\begingroup\@sanitize@url \@url }%
\providecommand \@url [1]{\endgroup\@href {#1}{\urlprefix }}%
\providecommand \urlprefix  [0]{URL }%
\providecommand \Eprint [0]{\href }%
\providecommand \doibase [0]{http://dx.doi.org/}%
\providecommand \selectlanguage [0]{\@gobble}%
\providecommand \bibinfo  [0]{\@secondoftwo}%
\providecommand \bibfield  [0]{\@secondoftwo}%
\providecommand \translation [1]{[#1]}%
\providecommand \BibitemOpen [0]{}%
\providecommand \bibitemStop [0]{}%
\providecommand \bibitemNoStop [0]{.\EOS\space}%
\providecommand \EOS [0]{\spacefactor3000\relax}%
\providecommand \BibitemShut  [1]{\csname bibitem#1\endcsname}%
\let\auto@bib@innerbib\@empty
%</preamble>
\bibitem [{\citenamefont {Soykal}\ and\ \citenamefont
  {Flatt\'{e}}(2010)}]{Soykal:prl:2010}%
  \BibitemOpen
  \bibfield  {author} {\bibinfo {author} {\bibfnamefont {O.~O.}\ \bibnamefont
  {Soykal}}\ and\ \bibinfo {author} {\bibfnamefont {M.~E.}\ \bibnamefont
  {Flatt\'{e}}},\ }\href {\doibase 10.1103/PhysRevLett.104.077202} {\bibfield
  {journal} {\bibinfo  {journal} {Phys. Rev. Lett.}\ }\textbf {\bibinfo
  {volume} {104}},\ \bibinfo {pages} {077202} (\bibinfo {year}
  {2010})}\BibitemShut {NoStop}%
\bibitem [{\citenamefont {Huebl}\ \emph {et~al.}(2013)\citenamefont {Huebl},
  \citenamefont {Zollitsch}, \citenamefont {Lotze}, \citenamefont {Hocke},
  \citenamefont {Greifenstein}, \citenamefont {Marx}, \citenamefont {Gross},\
  and\ \citenamefont {Goennenwein}}]{Huebl:prl:2013}%
  \BibitemOpen
  \bibfield  {author} {\bibinfo {author} {\bibfnamefont {H.}~\bibnamefont
  {Huebl}}, \bibinfo {author} {\bibfnamefont {C.~W.}\ \bibnamefont
  {Zollitsch}}, \bibinfo {author} {\bibfnamefont {J.}~\bibnamefont {Lotze}},
  \bibinfo {author} {\bibfnamefont {F.}~\bibnamefont {Hocke}}, \bibinfo
  {author} {\bibfnamefont {M.}~\bibnamefont {Greifenstein}}, \bibinfo {author}
  {\bibfnamefont {A.}~\bibnamefont {Marx}}, \bibinfo {author} {\bibfnamefont
  {R.}~\bibnamefont {Gross}}, \ and\ \bibinfo {author} {\bibfnamefont
  {S.~T.~B.}\ \bibnamefont {Goennenwein}},\ }\href {\doibase
  10.1103/PhysRevLett.111.127003} {\bibfield  {journal} {\bibinfo  {journal}
  {Phys. Rev. Lett.}\ }\textbf {\bibinfo {volume} {111}},\ \bibinfo {pages}
  {127003} (\bibinfo {year} {2013})}\BibitemShut {NoStop}%
\bibitem [{\citenamefont {Tabuchi}\ \emph {et~al.}(2014)\citenamefont
  {Tabuchi}, \citenamefont {Ishino}, \citenamefont {Ishikawa}, \citenamefont
  {Yamazaki}, \citenamefont {Usami},\ and\ \citenamefont
  {Nakamura}}]{Tabuchi:prl:2014}%
  \BibitemOpen
  \bibfield  {author} {\bibinfo {author} {\bibfnamefont {Y.}~\bibnamefont
  {Tabuchi}}, \bibinfo {author} {\bibfnamefont {S.}~\bibnamefont {Ishino}},
  \bibinfo {author} {\bibfnamefont {T.}~\bibnamefont {Ishikawa}}, \bibinfo
  {author} {\bibfnamefont {R.}~\bibnamefont {Yamazaki}}, \bibinfo {author}
  {\bibfnamefont {K.}~\bibnamefont {Usami}}, \ and\ \bibinfo {author}
  {\bibfnamefont {Y.}~\bibnamefont {Nakamura}},\ }\href {\doibase
  10.1103/PhysRevLett.113.083603} {\bibfield  {journal} {\bibinfo  {journal}
  {Phys. Rev. Lett.}\ }\textbf {\bibinfo {volume} {113}},\ \bibinfo {pages}
  {083603} (\bibinfo {year} {2014})}\BibitemShut {NoStop}%
\bibitem [{\citenamefont {Osada}\ \emph {et~al.}(2016)\citenamefont {Osada},
  \citenamefont {Hisatomi}, \citenamefont {Noguchi}, \citenamefont {Tabuchi},
  \citenamefont {Yamazaki}, \citenamefont {Usami}, \citenamefont {Sadgrove},
  \citenamefont {Yalla}, \citenamefont {Nomura},\ and\ \citenamefont
  {Nakamura}}]{Osada:prl:2016}%
  \BibitemOpen
  \bibfield  {author} {\bibinfo {author} {\bibfnamefont {A.}~\bibnamefont
  {Osada}}, \bibinfo {author} {\bibfnamefont {R.}~\bibnamefont {Hisatomi}},
  \bibinfo {author} {\bibfnamefont {A.}~\bibnamefont {Noguchi}}, \bibinfo
  {author} {\bibfnamefont {Y.}~\bibnamefont {Tabuchi}}, \bibinfo {author}
  {\bibfnamefont {R.}~\bibnamefont {Yamazaki}}, \bibinfo {author}
  {\bibfnamefont {K.}~\bibnamefont {Usami}}, \bibinfo {author} {\bibfnamefont
  {M.}~\bibnamefont {Sadgrove}}, \bibinfo {author} {\bibfnamefont
  {R.}~\bibnamefont {Yalla}}, \bibinfo {author} {\bibfnamefont
  {M.}~\bibnamefont {Nomura}}, \ and\ \bibinfo {author} {\bibfnamefont
  {Y.}~\bibnamefont {Nakamura}},\ }\href {\doibase
  10.1103/PhysRevLett.116.223601} {\bibfield  {journal} {\bibinfo  {journal}
  {Phys. Rev. Lett.}\ }\textbf {\bibinfo {volume} {116}},\ \bibinfo {pages}
  {223601} (\bibinfo {year} {2016})}\BibitemShut {NoStop}%
\bibitem [{\citenamefont {Zhang}\ \emph {et~al.}(2016)\citenamefont {Zhang},
  \citenamefont {Zhu}, \citenamefont {Zou},\ and\ \citenamefont
  {Tang}}]{Zhang:prl:2016}%
  \BibitemOpen
  \bibfield  {author} {\bibinfo {author} {\bibfnamefont {X.}~\bibnamefont
  {Zhang}}, \bibinfo {author} {\bibfnamefont {N.}~\bibnamefont {Zhu}}, \bibinfo
  {author} {\bibfnamefont {C.-L.}\ \bibnamefont {Zou}}, \ and\ \bibinfo
  {author} {\bibfnamefont {H.~X.}\ \bibnamefont {Tang}},\ }\href {\doibase
  10.1103/PhysRevLett.117.123605} {\bibfield  {journal} {\bibinfo  {journal}
  {Phys. Rev. Lett.}\ }\textbf {\bibinfo {volume} {117}},\ \bibinfo {pages}
  {123605} (\bibinfo {year} {2016})}\BibitemShut {NoStop}%
\bibitem [{\citenamefont {Haigh}\ \emph {et~al.}(2016)\citenamefont {Haigh},
  \citenamefont {Nunnenkamp}, \citenamefont {Ramsay},\ and\ \citenamefont
  {Ferguson}}]{Haigh:prl:2016}%
  \BibitemOpen
  \bibfield  {author} {\bibinfo {author} {\bibfnamefont {J.~A.}\ \bibnamefont
  {Haigh}}, \bibinfo {author} {\bibfnamefont {A.}~\bibnamefont {Nunnenkamp}},
  \bibinfo {author} {\bibfnamefont {A.~J.}\ \bibnamefont {Ramsay}}, \ and\
  \bibinfo {author} {\bibfnamefont {A.~J.}\ \bibnamefont {Ferguson}},\ }\href
  {\doibase 10.1103/PhysRevLett.117.133602} {\bibfield  {journal} {\bibinfo
  {journal} {Phys. Rev. Lett.}\ }\textbf {\bibinfo {volume} {117}},\ \bibinfo
  {pages} {133602} (\bibinfo {year} {2016})}\BibitemShut {NoStop}%
\bibitem [{\citenamefont {Braggio}\ \emph {et~al.}(2017)\citenamefont
  {Braggio}, \citenamefont {Carugno}, \citenamefont {Guarise}, \citenamefont
  {Ortolan},\ and\ \citenamefont {Ruoso}}]{Braggio:prl:2017}%
  \BibitemOpen
  \bibfield  {author} {\bibinfo {author} {\bibfnamefont {C.}~\bibnamefont
  {Braggio}}, \bibinfo {author} {\bibfnamefont {G.}~\bibnamefont {Carugno}},
  \bibinfo {author} {\bibfnamefont {M.}~\bibnamefont {Guarise}}, \bibinfo
  {author} {\bibfnamefont {A.}~\bibnamefont {Ortolan}}, \ and\ \bibinfo
  {author} {\bibfnamefont {G.}~\bibnamefont {Ruoso}},\ }\href {\doibase
  10.1103/PhysRevLett.118.107205} {\bibfield  {journal} {\bibinfo  {journal}
  {Phys. Rev. Lett.}\ }\textbf {\bibinfo {volume} {118}},\ \bibinfo {pages}
  {107205} (\bibinfo {year} {2017})}\BibitemShut {NoStop}%
\bibitem [{\citenamefont {Zhang}\ \emph {et~al.}(2014)\citenamefont {Zhang},
  \citenamefont {Zou}, \citenamefont {Jiang},\ and\ \citenamefont
  {Tang}}]{Zhang:prl:2014}%
  \BibitemOpen
  \bibfield  {author} {\bibinfo {author} {\bibfnamefont {X.}~\bibnamefont
  {Zhang}}, \bibinfo {author} {\bibfnamefont {C.-L.}\ \bibnamefont {Zou}},
  \bibinfo {author} {\bibfnamefont {L.}~\bibnamefont {Jiang}}, \ and\ \bibinfo
  {author} {\bibfnamefont {H.~X.}\ \bibnamefont {Tang}},\ }\href {\doibase
  10.1103/PhysRevLett.113.156401} {\bibfield  {journal} {\bibinfo  {journal}
  {Phys. Rev. Lett.}\ }\textbf {\bibinfo {volume} {113}},\ \bibinfo {pages}
  {156401} (\bibinfo {year} {2014})}\BibitemShut {NoStop}%
\bibitem [{\citenamefont {Goryachev}\ \emph {et~al.}(2014)\citenamefont
  {Goryachev}, \citenamefont {Farr}, \citenamefont {Creedon}, \citenamefont
  {Fan}, \citenamefont {Kostylev},\ and\ \citenamefont
  {Tobar}}]{Goryachev:pra:2014}%
  \BibitemOpen
  \bibfield  {author} {\bibinfo {author} {\bibfnamefont {M.}~\bibnamefont
  {Goryachev}}, \bibinfo {author} {\bibfnamefont {W.~G.}\ \bibnamefont {Farr}},
  \bibinfo {author} {\bibfnamefont {D.~L.}\ \bibnamefont {Creedon}}, \bibinfo
  {author} {\bibfnamefont {Y.}~\bibnamefont {Fan}}, \bibinfo {author}
  {\bibfnamefont {M.}~\bibnamefont {Kostylev}}, \ and\ \bibinfo {author}
  {\bibfnamefont {M.~E.}\ \bibnamefont {Tobar}},\ }\href {\doibase
  10.1103/PhysRevApplied.2.054002} {\bibfield  {journal} {\bibinfo  {journal}
  {Phys. Rev. Applied}\ }\textbf {\bibinfo {volume} {2}},\ \bibinfo {pages}
  {054002} (\bibinfo {year} {2014})}\BibitemShut {NoStop}%
\bibitem [{\citenamefont {Bourhill}\ \emph {et~al.}(2016)\citenamefont
  {Bourhill}, \citenamefont {Kostylev}, \citenamefont {Goryachev},
  \citenamefont {Creedon},\ and\ \citenamefont {Tobar}}]{Bourhill:prb:2016}%
  \BibitemOpen
  \bibfield  {author} {\bibinfo {author} {\bibfnamefont {J.}~\bibnamefont
  {Bourhill}}, \bibinfo {author} {\bibfnamefont {N.}~\bibnamefont {Kostylev}},
  \bibinfo {author} {\bibfnamefont {M.}~\bibnamefont {Goryachev}}, \bibinfo
  {author} {\bibfnamefont {D.~L.}\ \bibnamefont {Creedon}}, \ and\ \bibinfo
  {author} {\bibfnamefont {M.~E.}\ \bibnamefont {Tobar}},\ }\href {\doibase
  10.1103/PhysRevB.93.144420} {\bibfield  {journal} {\bibinfo  {journal} {Phys.
  Rev. B}\ }\textbf {\bibinfo {volume} {93}},\ \bibinfo {pages} {144420}
  (\bibinfo {year} {2016})}\BibitemShut {NoStop}%
\bibitem [{\citenamefont {Kostylev}\ \emph {et~al.}(2016)\citenamefont
  {Kostylev}, \citenamefont {Goryachev},\ and\ \citenamefont
  {Tobar}}]{Kostylev:apl:2016}%
  \BibitemOpen
  \bibfield  {author} {\bibinfo {author} {\bibfnamefont {N.}~\bibnamefont
  {Kostylev}}, \bibinfo {author} {\bibfnamefont {M.}~\bibnamefont {Goryachev}},
  \ and\ \bibinfo {author} {\bibfnamefont {M.~E.}\ \bibnamefont {Tobar}},\
  }\href {\doibase 10.1063/1.4941730} {\bibfield  {journal} {\bibinfo
  {journal} {Appl. Phys. Lett.}\ }\textbf {\bibinfo {volume} {108}},\ \bibinfo
  {pages} {062402} (\bibinfo {year} {2016})}\BibitemShut {NoStop}%
\bibitem [{\citenamefont {Maier-Flaig}\ \emph {et~al.}(2017)\citenamefont
  {Maier-Flaig}, \citenamefont {Harder}, \citenamefont {Klingler},
  \citenamefont {Qiu}, \citenamefont {Saitoh}, \citenamefont {Weiler},
  \citenamefont {Geprägs}, \citenamefont {Gross}, \citenamefont
  {Goennenwein},\ and\ \citenamefont {Huebl}}]{Maier-Flaig:apl:2017}%
  \BibitemOpen
  \bibfield  {author} {\bibinfo {author} {\bibfnamefont {H.}~\bibnamefont
  {Maier-Flaig}}, \bibinfo {author} {\bibfnamefont {M.}~\bibnamefont {Harder}},
  \bibinfo {author} {\bibfnamefont {S.}~\bibnamefont {Klingler}}, \bibinfo
  {author} {\bibfnamefont {Z.}~\bibnamefont {Qiu}}, \bibinfo {author}
  {\bibfnamefont {E.}~\bibnamefont {Saitoh}}, \bibinfo {author} {\bibfnamefont
  {M.}~\bibnamefont {Weiler}}, \bibinfo {author} {\bibfnamefont
  {S.}~\bibnamefont {Geprägs}}, \bibinfo {author} {\bibfnamefont
  {R.}~\bibnamefont {Gross}}, \bibinfo {author} {\bibfnamefont {S.~T.~B.}\
  \bibnamefont {Goennenwein}}, \ and\ \bibinfo {author} {\bibfnamefont
  {H.}~\bibnamefont {Huebl}},\ }\href {\doibase 10.1063/1.4979409} {\bibfield
  {journal} {\bibinfo  {journal} {Appl. Phys. Lett.}\ }\textbf {\bibinfo
  {volume} {110}},\ \bibinfo {pages} {132401} (\bibinfo {year}
  {2017})}\BibitemShut {NoStop}%
\bibitem [{\citenamefont {Boventer}\ \emph {et~al.}(2018)\citenamefont
  {Boventer}, \citenamefont {Pfirrmann}, \citenamefont {Krause}, \citenamefont
  {{Sch\"{o}n}}, \citenamefont {{Kl\"{a}ui}},\ and\ \citenamefont
  {Weides}}]{Boventer2018}%
  \BibitemOpen
  \bibfield  {author} {\bibinfo {author} {\bibfnamefont {I.}~\bibnamefont
  {Boventer}}, \bibinfo {author} {\bibfnamefont {M.}~\bibnamefont {Pfirrmann}},
  \bibinfo {author} {\bibfnamefont {J.}~\bibnamefont {Krause}}, \bibinfo
  {author} {\bibfnamefont {Y.}~\bibnamefont {{Sch\"{o}n}}}, \bibinfo {author}
  {\bibfnamefont {M.}~\bibnamefont {{Kl\"{a}ui}}}, \ and\ \bibinfo {author}
  {\bibfnamefont {M.}~\bibnamefont {Weides}},\ }\href
  {https://arxiv.org/abs/1801.01439} {\bibfield  {journal} {\bibinfo  {journal}
  {arXiv}\ } (\bibinfo {year} {2018})},\ \Eprint
  {http://arxiv.org/abs/1801.01439} {arXiv:1801.01439} \BibitemShut {NoStop}%
\bibitem [{\citenamefont {Klingler}\ \emph {et~al.}(2016)\citenamefont
  {Klingler}, \citenamefont {Maier-Flaig}, \citenamefont {Gross}, \citenamefont
  {Hu}, \citenamefont {Huebl}, \citenamefont {Goennenwein},\ and\ \citenamefont
  {Weiler}}]{Klingler:apl:2016}%
  \BibitemOpen
  \bibfield  {author} {\bibinfo {author} {\bibfnamefont {S.}~\bibnamefont
  {Klingler}}, \bibinfo {author} {\bibfnamefont {H.}~\bibnamefont
  {Maier-Flaig}}, \bibinfo {author} {\bibfnamefont {R.}~\bibnamefont {Gross}},
  \bibinfo {author} {\bibfnamefont {C.-M.}\ \bibnamefont {Hu}}, \bibinfo
  {author} {\bibfnamefont {H.}~\bibnamefont {Huebl}}, \bibinfo {author}
  {\bibfnamefont {S.~T.~B.}\ \bibnamefont {Goennenwein}}, \ and\ \bibinfo
  {author} {\bibfnamefont {M.}~\bibnamefont {Weiler}},\ }\href {\doibase
  10.1063/1.4961052} {\bibfield  {journal} {\bibinfo  {journal} {Appl. Phys.
  Lett.}\ }\textbf {\bibinfo {volume} {109}},\ \bibinfo {pages} {072402}
  (\bibinfo {year} {2016})}\BibitemShut {NoStop}%
\bibitem [{\citenamefont {Cao}\ \emph {et~al.}(2015)\citenamefont {Cao},
  \citenamefont {Yan}, \citenamefont {Huebl}, \citenamefont {Goennenwein},\
  and\ \citenamefont {Bauer}}]{Cao:prb:2015}%
  \BibitemOpen
  \bibfield  {author} {\bibinfo {author} {\bibfnamefont {Y.}~\bibnamefont
  {Cao}}, \bibinfo {author} {\bibfnamefont {P.}~\bibnamefont {Yan}}, \bibinfo
  {author} {\bibfnamefont {H.}~\bibnamefont {Huebl}}, \bibinfo {author}
  {\bibfnamefont {S.~T.~B.}\ \bibnamefont {Goennenwein}}, \ and\ \bibinfo
  {author} {\bibfnamefont {G.~E.~W.}\ \bibnamefont {Bauer}},\ }\href {\doibase
  10.1103/PhysRevB.91.094423} {\bibfield  {journal} {\bibinfo  {journal} {Phys.
  Rev. B}\ }\textbf {\bibinfo {volume} {91}},\ \bibinfo {pages} {094423}
  (\bibinfo {year} {2015})}\BibitemShut {NoStop}%
\bibitem [{\citenamefont {Bai}\ \emph {et~al.}(2015)\citenamefont {Bai},
  \citenamefont {Harder}, \citenamefont {Chen}, \citenamefont {Fan},
  \citenamefont {Xiao},\ and\ \citenamefont {Hu}}]{Bai:prl:2015}%
  \BibitemOpen
  \bibfield  {author} {\bibinfo {author} {\bibfnamefont {L.}~\bibnamefont
  {Bai}}, \bibinfo {author} {\bibfnamefont {M.}~\bibnamefont {Harder}},
  \bibinfo {author} {\bibfnamefont {Y.~P.}\ \bibnamefont {Chen}}, \bibinfo
  {author} {\bibfnamefont {X.}~\bibnamefont {Fan}}, \bibinfo {author}
  {\bibfnamefont {J.~Q.}\ \bibnamefont {Xiao}}, \ and\ \bibinfo {author}
  {\bibfnamefont {C.-M.}\ \bibnamefont {Hu}},\ }\href {\doibase
  10.1103/PhysRevLett.114.227201} {\bibfield  {journal} {\bibinfo  {journal}
  {Phys. Rev. Lett.}\ }\textbf {\bibinfo {volume} {114}},\ \bibinfo {pages}
  {227201} (\bibinfo {year} {2015})}\BibitemShut {NoStop}%
\bibitem [{\citenamefont {Bai}\ \emph {et~al.}(2017)\citenamefont {Bai},
  \citenamefont {Harder}, \citenamefont {Hyde}, \citenamefont {Zhang},
  \citenamefont {Hu}, \citenamefont {Chen},\ and\ \citenamefont
  {Xiao}}]{Bai:prl:2017}%
  \BibitemOpen
  \bibfield  {author} {\bibinfo {author} {\bibfnamefont {L.}~\bibnamefont
  {Bai}}, \bibinfo {author} {\bibfnamefont {M.}~\bibnamefont {Harder}},
  \bibinfo {author} {\bibfnamefont {P.}~\bibnamefont {Hyde}}, \bibinfo {author}
  {\bibfnamefont {Z.}~\bibnamefont {Zhang}}, \bibinfo {author} {\bibfnamefont
  {C.-M.}\ \bibnamefont {Hu}}, \bibinfo {author} {\bibfnamefont {Y.~P.}\
  \bibnamefont {Chen}}, \ and\ \bibinfo {author} {\bibfnamefont {J.~Q.}\
  \bibnamefont {Xiao}},\ }\href {\doibase 10.1103/PhysRevLett.118.217201}
  {\bibfield  {journal} {\bibinfo  {journal} {Phys. Rev. Lett.}\ }\textbf
  {\bibinfo {volume} {118}},\ \bibinfo {pages} {217201} (\bibinfo {year}
  {2017})}\BibitemShut {NoStop}%
\bibitem [{\citenamefont {Yuan}\ and\ \citenamefont
  {Wang}(2017)}]{Yuan:apl:2017}%
  \BibitemOpen
  \bibfield  {author} {\bibinfo {author} {\bibfnamefont {H.~Y.}\ \bibnamefont
  {Yuan}}\ and\ \bibinfo {author} {\bibfnamefont {X.~R.}\ \bibnamefont
  {Wang}},\ }\href@noop {} {\bibfield  {journal} {\bibinfo  {journal} {Appl.
  Phys. Lett.}\ }\textbf {\bibinfo {volume} {110}},\ \bibinfo {pages} {082403}
  (\bibinfo {year} {2017})}\BibitemShut {NoStop}%
\bibitem [{\citenamefont {Mergenthaler}\ \emph {et~al.}(2017)\citenamefont
  {Mergenthaler}, \citenamefont {Liu}, \citenamefont {Le~Roy}, \citenamefont
  {Ares}, \citenamefont {Thompson}, \citenamefont {Bogani}, \citenamefont
  {Luis}, \citenamefont {Blundell}, \citenamefont {Lancaster}, \citenamefont
  {Ardavan}, \citenamefont {Briggs}, \citenamefont {Leek},\ and\ \citenamefont
  {Laird}}]{Mergenthaler:prl:2017}%
  \BibitemOpen
  \bibfield  {author} {\bibinfo {author} {\bibfnamefont {M.}~\bibnamefont
  {Mergenthaler}}, \bibinfo {author} {\bibfnamefont {J.}~\bibnamefont {Liu}},
  \bibinfo {author} {\bibfnamefont {J.~J.}\ \bibnamefont {Le~Roy}}, \bibinfo
  {author} {\bibfnamefont {N.}~\bibnamefont {Ares}}, \bibinfo {author}
  {\bibfnamefont {A.~L.}\ \bibnamefont {Thompson}}, \bibinfo {author}
  {\bibfnamefont {L.}~\bibnamefont {Bogani}}, \bibinfo {author} {\bibfnamefont
  {F.}~\bibnamefont {Luis}}, \bibinfo {author} {\bibfnamefont {S.~J.}\
  \bibnamefont {Blundell}}, \bibinfo {author} {\bibfnamefont {T.}~\bibnamefont
  {Lancaster}}, \bibinfo {author} {\bibfnamefont {A.}~\bibnamefont {Ardavan}},
  \bibinfo {author} {\bibfnamefont {G.~A.~D.}\ \bibnamefont {Briggs}}, \bibinfo
  {author} {\bibfnamefont {P.~J.}\ \bibnamefont {Leek}}, \ and\ \bibinfo
  {author} {\bibfnamefont {E.~A.}\ \bibnamefont {Laird}},\ }\href {\doibase
  10.1103/PhysRevLett.119.147701} {\bibfield  {journal} {\bibinfo  {journal}
  {Phys. Rev. Lett.}\ }\textbf {\bibinfo {volume} {119}},\ \bibinfo {pages}
  {147701} (\bibinfo {year} {2017})}\BibitemShut {NoStop}%
\bibitem [{\citenamefont {Lambert}\ \emph {et~al.}(2016)\citenamefont
  {Lambert}, \citenamefont {Haigh}, \citenamefont {Langenfeld}, \citenamefont
  {Doherty},\ and\ \citenamefont {Ferguson}}]{Lambert:pra:2016}%
  \BibitemOpen
  \bibfield  {author} {\bibinfo {author} {\bibfnamefont {N.~J.}\ \bibnamefont
  {Lambert}}, \bibinfo {author} {\bibfnamefont {J.~A.}\ \bibnamefont {Haigh}},
  \bibinfo {author} {\bibfnamefont {S.}~\bibnamefont {Langenfeld}}, \bibinfo
  {author} {\bibfnamefont {A.~C.}\ \bibnamefont {Doherty}}, \ and\ \bibinfo
  {author} {\bibfnamefont {A.~J.}\ \bibnamefont {Ferguson}},\ }\href {\doibase
  10.1103/PhysRevA.93.021803} {\bibfield  {journal} {\bibinfo  {journal} {Phys.
  Rev. A}\ }\textbf {\bibinfo {volume} {93}},\ \bibinfo {pages} {021803}
  (\bibinfo {year} {2016})}\BibitemShut {NoStop}%
\bibitem [{\citenamefont {Zare~Rameshti}\ and\ \citenamefont
  {Bauer}(2018)}]{Rameshti:prb:2018}%
  \BibitemOpen
  \bibfield  {author} {\bibinfo {author} {\bibfnamefont {B.}~\bibnamefont
  {Zare~Rameshti}}\ and\ \bibinfo {author} {\bibfnamefont {G.~E.~W.}\
  \bibnamefont {Bauer}},\ }\href {\doibase 10.1103/PhysRevB.97.014419}
  {\bibfield  {journal} {\bibinfo  {journal} {Phys. Rev. B}\ }\textbf {\bibinfo
  {volume} {97}},\ \bibinfo {pages} {014419} (\bibinfo {year}
  {2018})}\BibitemShut {NoStop}%
\bibitem [{\citenamefont {Cheng}\ \emph {et~al.}(2014)\citenamefont {Cheng},
  \citenamefont {Xiao}, \citenamefont {Niu},\ and\ \citenamefont
  {Brataas}}]{Cheng:prl2014}%
  \BibitemOpen
  \bibfield  {author} {\bibinfo {author} {\bibfnamefont {R.}~\bibnamefont
  {Cheng}}, \bibinfo {author} {\bibfnamefont {J.}~\bibnamefont {Xiao}},
  \bibinfo {author} {\bibfnamefont {Q.}~\bibnamefont {Niu}}, \ and\ \bibinfo
  {author} {\bibfnamefont {A.}~\bibnamefont {Brataas}},\ }\href@noop {}
  {\bibfield  {journal} {\bibinfo  {journal} {Phys. Rev. Lett.}\ }\textbf
  {\bibinfo {volume} {113}},\ \bibinfo {pages} {057601} (\bibinfo {year}
  {2014})}\BibitemShut {NoStop}%
\bibitem [{\citenamefont {Marti}\ \emph {et~al.}(2014)\citenamefont {Marti},
  \citenamefont {Fina}, \citenamefont {Frontera}, \citenamefont {Liu},
  \citenamefont {Wadley}, \citenamefont {He}, \citenamefont {Paull},
  \citenamefont {Clarkson}, \citenamefont {Kudrnovsk{\'y}}, \citenamefont
  {Turek}, \citenamefont {Kunes}, \citenamefont {Yi}, \citenamefont {Chu},
  \citenamefont {Nelson}, \citenamefont {You}, \citenamefont {Arenholz},
  \citenamefont {Salahuddin}, \citenamefont {Fontcuberta}, \citenamefont
  {Jungwirth},\ and\ \citenamefont {Ramesh}}]{Marti2014}%
  \BibitemOpen
  \bibfield  {author} {\bibinfo {author} {\bibfnamefont {X.}~\bibnamefont
  {Marti}}, \bibinfo {author} {\bibfnamefont {I.}~\bibnamefont {Fina}},
  \bibinfo {author} {\bibfnamefont {C.}~\bibnamefont {Frontera}}, \bibinfo
  {author} {\bibfnamefont {J.}~\bibnamefont {Liu}}, \bibinfo {author}
  {\bibfnamefont {P.}~\bibnamefont {Wadley}}, \bibinfo {author} {\bibfnamefont
  {Q.}~\bibnamefont {He}}, \bibinfo {author} {\bibfnamefont {R.~J.}\
  \bibnamefont {Paull}}, \bibinfo {author} {\bibfnamefont {J.~D.}\ \bibnamefont
  {Clarkson}}, \bibinfo {author} {\bibfnamefont {J.}~\bibnamefont
  {Kudrnovsk{\'y}}}, \bibinfo {author} {\bibfnamefont {I.}~\bibnamefont
  {Turek}}, \bibinfo {author} {\bibfnamefont {J.}~\bibnamefont {Kunes}},
  \bibinfo {author} {\bibfnamefont {D.}~\bibnamefont {Yi}}, \bibinfo {author}
  {\bibfnamefont {J.-H.}\ \bibnamefont {Chu}}, \bibinfo {author} {\bibfnamefont
  {C.~T.}\ \bibnamefont {Nelson}}, \bibinfo {author} {\bibfnamefont
  {L.}~\bibnamefont {You}}, \bibinfo {author} {\bibfnamefont {E.}~\bibnamefont
  {Arenholz}}, \bibinfo {author} {\bibfnamefont {S.}~\bibnamefont
  {Salahuddin}}, \bibinfo {author} {\bibfnamefont {J.}~\bibnamefont
  {Fontcuberta}}, \bibinfo {author} {\bibfnamefont {T.}~\bibnamefont
  {Jungwirth}}, \ and\ \bibinfo {author} {\bibfnamefont {R.}~\bibnamefont
  {Ramesh}},\ }\href@noop {} {\bibfield  {journal} {\bibinfo  {journal} {Nat.
  Mater.}\ }\textbf {\bibinfo {volume} {13}},\ \bibinfo {pages} {367 EP }
  (\bibinfo {year} {2014})}\BibitemShut {NoStop}%
\bibitem [{\citenamefont {Cheng}\ \emph {et~al.}(2016)\citenamefont {Cheng},
  \citenamefont {Xiao},\ and\ \citenamefont {Brataas}}]{Cheng:prl2016}%
  \BibitemOpen
  \bibfield  {author} {\bibinfo {author} {\bibfnamefont {R.}~\bibnamefont
  {Cheng}}, \bibinfo {author} {\bibfnamefont {D.}~\bibnamefont {Xiao}}, \ and\
  \bibinfo {author} {\bibfnamefont {A.}~\bibnamefont {Brataas}},\ }\href@noop
  {} {\bibfield  {journal} {\bibinfo  {journal} {Phys. Rev. Lett.}\ }\textbf
  {\bibinfo {volume} {116}},\ \bibinfo {pages} {207603} (\bibinfo {year}
  {2016})}\BibitemShut {NoStop}%
\bibitem [{\citenamefont {Jungwirth}\ \emph {et~al.}(2016)\citenamefont
  {Jungwirth}, \citenamefont {Marti}, \citenamefont {Wadley},\ and\
  \citenamefont {Wunderlich}}]{Jungwirth2016}%
  \BibitemOpen
  \bibfield  {author} {\bibinfo {author} {\bibfnamefont {T.}~\bibnamefont
  {Jungwirth}}, \bibinfo {author} {\bibfnamefont {X.}~\bibnamefont {Marti}},
  \bibinfo {author} {\bibfnamefont {P.}~\bibnamefont {Wadley}}, \ and\ \bibinfo
  {author} {\bibfnamefont {J.}~\bibnamefont {Wunderlich}},\ }\href
  {http://dx.doi.org/10.1038/nnano.2016.18} {\bibfield  {journal} {\bibinfo
  {journal} {Nat. Nanotechnol.}\ }\textbf {\bibinfo {volume} {11}},\ \bibinfo
  {pages} {231 EP } (\bibinfo {year} {2016})}\BibitemShut {NoStop}%
\bibitem [{\citenamefont {Baltz}\ \emph {et~al.}(2018)\citenamefont {Baltz},
  \citenamefont {Manchon}, \citenamefont {Tsoi}, \citenamefont {Moriyama},
  \citenamefont {Ono},\ and\ \citenamefont {Tserkovnyak}}]{Baltz2018}%
  \BibitemOpen
  \bibfield  {author} {\bibinfo {author} {\bibfnamefont {V.}~\bibnamefont
  {Baltz}}, \bibinfo {author} {\bibfnamefont {A.}~\bibnamefont {Manchon}},
  \bibinfo {author} {\bibfnamefont {M.}~\bibnamefont {Tsoi}}, \bibinfo {author}
  {\bibfnamefont {T.}~\bibnamefont {Moriyama}}, \bibinfo {author}
  {\bibfnamefont {T.}~\bibnamefont {Ono}}, \ and\ \bibinfo {author}
  {\bibfnamefont {Y.}~\bibnamefont {Tserkovnyak}},\ }\href {\doibase
  10.1103/RevModPhys.90.015005} {\bibfield  {journal} {\bibinfo  {journal}
  {Rev. Mod. Phys.}\ }\textbf {\bibinfo {volume} {90}},\ \bibinfo {pages}
  {015005} (\bibinfo {year} {2018})}\BibitemShut {NoStop}%
\bibitem [{\citenamefont {Holstein}\ and\ \citenamefont
  {Primakoff}(1940)}]{HolsteinPrimakoff}%
  \BibitemOpen
  \bibfield  {author} {\bibinfo {author} {\bibfnamefont {T.}~\bibnamefont
  {Holstein}}\ and\ \bibinfo {author} {\bibfnamefont {H.}~\bibnamefont
  {Primakoff}},\ }\href {\doibase 10.1103/PhysRev.58.1098} {\bibfield
  {journal} {\bibinfo  {journal} {Phys. Rev.}\ }\textbf {\bibinfo {volume}
  {58}},\ \bibinfo {pages} {1098} (\bibinfo {year} {1940})}\BibitemShut
  {NoStop}%
\bibitem [{\citenamefont {Kamra}\ \emph {et~al.}(2017)\citenamefont {Kamra},
  \citenamefont {Agrawal},\ and\ \citenamefont {Belzig}}]{Kamra:prb:2017}%
  \BibitemOpen
  \bibfield  {author} {\bibinfo {author} {\bibfnamefont {A.}~\bibnamefont
  {Kamra}}, \bibinfo {author} {\bibfnamefont {U.}~\bibnamefont {Agrawal}}, \
  and\ \bibinfo {author} {\bibfnamefont {W.}~\bibnamefont {Belzig}},\ }\href
  {\doibase 10.1103/PhysRevB.96.020411} {\bibfield  {journal} {\bibinfo
  {journal} {Phys. Rev. B}\ }\textbf {\bibinfo {volume} {96}},\ \bibinfo
  {pages} {020411} (\bibinfo {year} {2017})}\BibitemShut {NoStop}%
\bibitem [{\citenamefont {Kamra}\ and\ \citenamefont
  {Belzig}(2016)}]{Kamra:prl:2016}%
  \BibitemOpen
  \bibfield  {author} {\bibinfo {author} {\bibfnamefont {A.}~\bibnamefont
  {Kamra}}\ and\ \bibinfo {author} {\bibfnamefont {W.}~\bibnamefont {Belzig}},\
  }\href {\doibase 10.1103/PhysRevLett.116.146601} {\bibfield  {journal}
  {\bibinfo  {journal} {Phys. Rev. Lett.}\ }\textbf {\bibinfo {volume} {116}},\
  \bibinfo {pages} {146601} (\bibinfo {year} {2016})}\BibitemShut {NoStop}%
\bibitem [{\citenamefont {Satoh}\ \emph {et~al.}(2010)\citenamefont {Satoh},
  \citenamefont {Cho}, \citenamefont {Iida}, \citenamefont {Shimura},
  \citenamefont {Kuroda}, \citenamefont {Ueda}, \citenamefont {Ueda},
  \citenamefont {Ivanov}, \citenamefont {Nori},\ and\ \citenamefont
  {Fiebig}}]{Satoh:prl:2010}%
  \BibitemOpen
  \bibfield  {author} {\bibinfo {author} {\bibfnamefont {T.}~\bibnamefont
  {Satoh}}, \bibinfo {author} {\bibfnamefont {S.-J.}\ \bibnamefont {Cho}},
  \bibinfo {author} {\bibfnamefont {R.}~\bibnamefont {Iida}}, \bibinfo {author}
  {\bibfnamefont {T.}~\bibnamefont {Shimura}}, \bibinfo {author} {\bibfnamefont
  {K.}~\bibnamefont {Kuroda}}, \bibinfo {author} {\bibfnamefont
  {H.}~\bibnamefont {Ueda}}, \bibinfo {author} {\bibfnamefont {Y.}~\bibnamefont
  {Ueda}}, \bibinfo {author} {\bibfnamefont {B.~A.}\ \bibnamefont {Ivanov}},
  \bibinfo {author} {\bibfnamefont {F.}~\bibnamefont {Nori}}, \ and\ \bibinfo
  {author} {\bibfnamefont {M.}~\bibnamefont {Fiebig}},\ }\href {\doibase
  10.1103/PhysRevLett.105.077402} {\bibfield  {journal} {\bibinfo  {journal}
  {Phys. Rev. Lett.}\ }\textbf {\bibinfo {volume} {105}},\ \bibinfo {pages}
  {077402} (\bibinfo {year} {2010})}\BibitemShut {NoStop}%
\bibitem [{\citenamefont {Hutchings}\ and\ \citenamefont
  {Samuelsen}(1972)}]{Hutchings:prb:1972}%
  \BibitemOpen
  \bibfield  {author} {\bibinfo {author} {\bibfnamefont {M.~T.}\ \bibnamefont
  {Hutchings}}\ and\ \bibinfo {author} {\bibfnamefont {E.~J.}\ \bibnamefont
  {Samuelsen}},\ }\href {\doibase 10.1103/PhysRevB.6.3447} {\bibfield
  {journal} {\bibinfo  {journal} {Phys. Rev. B}\ }\textbf {\bibinfo {volume}
  {6}},\ \bibinfo {pages} {3447} (\bibinfo {year} {1972})}\BibitemShut
  {NoStop}%
\bibitem [{\citenamefont {Schoen}\ \emph {et~al.}(2016)\citenamefont {Schoen},
  \citenamefont {Thonig}, \citenamefont {Schneider}, \citenamefont {Silva},
  \citenamefont {Nembach}, \citenamefont {Eriksson}, \citenamefont {Karis},\
  and\ \citenamefont {Shaw}}]{Schoen2016}%
  \BibitemOpen
  \bibfield  {author} {\bibinfo {author} {\bibfnamefont {M.~A.~W.}\
  \bibnamefont {Schoen}}, \bibinfo {author} {\bibfnamefont {D.}~\bibnamefont
  {Thonig}}, \bibinfo {author} {\bibfnamefont {M.~L.}\ \bibnamefont
  {Schneider}}, \bibinfo {author} {\bibfnamefont {T.~J.}\ \bibnamefont
  {Silva}}, \bibinfo {author} {\bibfnamefont {H.~T.}\ \bibnamefont {Nembach}},
  \bibinfo {author} {\bibfnamefont {O.}~\bibnamefont {Eriksson}}, \bibinfo
  {author} {\bibfnamefont {O.}~\bibnamefont {Karis}}, \ and\ \bibinfo {author}
  {\bibfnamefont {J.~M.}\ \bibnamefont {Shaw}},\ }\href
  {http://dx.doi.org/10.1038/nphys3770} {\bibfield  {journal} {\bibinfo
  {journal} {Nat. Phys.}\ }\textbf {\bibinfo {volume} {12}},\ \bibinfo {pages}
  {839 EP } (\bibinfo {year} {2016})}\BibitemShut {NoStop}%
\bibitem [{\citenamefont {Gilleo}\ and\ \citenamefont
  {Geller}(1958)}]{YIGspindensity}%
  \BibitemOpen
  \bibfield  {author} {\bibinfo {author} {\bibfnamefont {M.~A.}\ \bibnamefont
  {Gilleo}}\ and\ \bibinfo {author} {\bibfnamefont {S.}~\bibnamefont
  {Geller}},\ }\href {\doibase 10.1103/PhysRev.110.73} {\bibfield  {journal}
  {\bibinfo  {journal} {Phys. Rev.}\ }\textbf {\bibinfo {volume} {110}},\
  \bibinfo {pages} {73} (\bibinfo {year} {1958})}\BibitemShut {NoStop}%
\bibitem [{\citenamefont {Hansen}\ \emph {et~al.}(1974)\citenamefont {Hansen},
  \citenamefont {Röschmann},\ and\ \citenamefont {Tolksdorf}}]{YIGMs}%
  \BibitemOpen
  \bibfield  {author} {\bibinfo {author} {\bibfnamefont {P.}~\bibnamefont
  {Hansen}}, \bibinfo {author} {\bibfnamefont {P.}~\bibnamefont {Röschmann}},
  \ and\ \bibinfo {author} {\bibfnamefont {W.}~\bibnamefont {Tolksdorf}},\
  }\href {\doibase 10.1063/1.1663657} {\bibfield  {journal} {\bibinfo
  {journal} {J. Appl. Phys.}\ }\textbf {\bibinfo {volume} {45}},\ \bibinfo
  {pages} {2728} (\bibinfo {year} {1974})}\BibitemShut {NoStop}%
\bibitem [{\citenamefont {Kotthaus}\ and\ \citenamefont
  {Jaccarino}(1972)}]{Kotthaus:prl:1972}%
  \BibitemOpen
  \bibfield  {author} {\bibinfo {author} {\bibfnamefont {J.~P.}\ \bibnamefont
  {Kotthaus}}\ and\ \bibinfo {author} {\bibfnamefont {V.}~\bibnamefont
  {Jaccarino}},\ }\href {\doibase 10.1103/PhysRevLett.28.1649} {\bibfield
  {journal} {\bibinfo  {journal} {Phys. Rev. Lett.}\ }\textbf {\bibinfo
  {volume} {28}},\ \bibinfo {pages} {1649} (\bibinfo {year}
  {1972})}\BibitemShut {NoStop}%
\bibitem [{\citenamefont {{Chappel, E.}}\ \emph {et~al.}(2000)\citenamefont
  {{Chappel, E.}}, \citenamefont {{Núñez-Regueiro, M. D.}}, \citenamefont
  {{Dupont, F.}}, \citenamefont {{Chouteau, G.}}, \citenamefont {{Darie, C.}},\
  and\ \citenamefont {{Sulpice, A.}}}]{NaNiO2}%
  \BibitemOpen
  \bibfield  {author} {\bibinfo {author} {\bibnamefont {{Chappel, E.}}},
  \bibinfo {author} {\bibnamefont {{Núñez-Regueiro, M. D.}}}, \bibinfo
  {author} {\bibnamefont {{Dupont, F.}}}, \bibinfo {author} {\bibnamefont
  {{Chouteau, G.}}}, \bibinfo {author} {\bibnamefont {{Darie, C.}}}, \ and\
  \bibinfo {author} {\bibnamefont {{Sulpice, A.}}},\ }\href {\doibase
  10.1007/s100510070098} {\bibfield  {journal} {\bibinfo  {journal} {Eur. Phys.
  J. B}\ }\textbf {\bibinfo {volume} {17}},\ \bibinfo {pages} {609} (\bibinfo
  {year} {2000})}\BibitemShut {NoStop}%
\bibitem [{\citenamefont {Schr\"{o}n}\ and\ \citenamefont
  {Bechstedt}(2015)}]{Schron:prb:2015}%
  \BibitemOpen
  \bibfield  {author} {\bibinfo {author} {\bibfnamefont {A.}~\bibnamefont
  {Schr\"{o}n}}\ and\ \bibinfo {author} {\bibfnamefont {F.}~\bibnamefont
  {Bechstedt}},\ }\href {\doibase 10.1103/PhysRevB.92.165112} {\bibfield
  {journal} {\bibinfo  {journal} {Phys. Rev. B}\ }\textbf {\bibinfo {volume}
  {92}},\ \bibinfo {pages} {165112} (\bibinfo {year} {2015})}\BibitemShut
  {NoStop}%
\bibitem [{\citenamefont {Ross}(2013)}]{Ross:TUMunchen:2013}%
  \BibitemOpen
  \bibfield  {author} {\bibinfo {author} {\bibfnamefont {M.~P.}\ \bibnamefont
  {Ross}},\ }\emph {\bibinfo {title} {Spin Dynamics in an Antiferromagnet}},\
  \href@noop {} {Ph.D. thesis},\ \bibinfo  {school} {Technische Universit\"{a}t
  M\"{u}nchen} (\bibinfo {year} {2013})\BibitemShut {NoStop}%
\end{thebibliography}
\end{document}